\documentclass[12pt]{article}

\usepackage{mathrsfs}
\usepackage{amsmath}
\usepackage{amssymb}
\usepackage{algorithmic}
\usepackage{algorithm}
\usepackage{graphicx}
\usepackage{multirow}
\usepackage{booktabs}
\usepackage{url}
\usepackage{caption}
\usepackage{xcolor}
\usepackage{geometry}
\usepackage{natbib}

\newcommand{\blind}{1}

\newtheorem{theorem}{Theorem}[section]

\newtheorem{remark}[theorem]{Remark}

\newtheorem{assumption}[theorem]{Assumption}

\newcommand{\bX}{{\mbox{\boldmath $X$}}}

\newcommand{\bF}{{\mbox{\boldmath $F$}}}

\newcommand{\bR}{{\mbox{\boldmath $R$}}}

\newcommand{\bD}{{\mbox{\boldmath $D$}}}
\newcommand{\bO}{{\mbox{\boldmath $O$}}}
\newcommand{\bQ}{{\mbox{\boldmath $Q$}}}

\newcommand{\bE}{{\mbox{\boldmath $E$}}}
\newcommand{\bU}{{\mbox{\boldmath $U$}}}
\newcommand{\bW}{{\mbox{\boldmath $W$}}}
\newcommand{\bL}{{\mbox{\boldmath $L$}}}

\newcommand{\bC}{{\mbox{\boldmath $C$}}}

\newcommand{\bbeta}{{\mbox{\boldmath $\beta$}}}
\newcommand{\balpha}{{\mbox{\boldmath $\alpha$}}}
\newcommand{\bepsilon}{{\mbox{\boldmath $\epsilon$}}}
\newcommand{\bbf}{{\mbox{\boldmath $f$}}}
\newcommand{\bSigma}{{\mbox{\boldmath $\Sigma$}}}
\newcommand{\bLambda}{{\mbox{\boldmath $\Lambda$}}}
\newcommand{\bGamma}{{\mbox{\boldmath $\Gamma$}}}

\newcommand{\bx}{{\mbox{\boldmath $x$}}}

\begin{document}

\def\spacingset#1{\renewcommand{\baselinestretch}%
{#1}\small\normalsize} \spacingset{1}

%%%%%%%%%%%%%%%%%%%%%%%%%%%%%%%%%%%%%%%%%%%%%%%%%%%%%%%%%%%%%%%%%%%%%%%%%%%%%%
%\title{ Matrix Quantile Factor Model}
%
%\author{Xin-Bing Kong
%\thanks{ School of Statistics and Data Science, Nanjing Audit University, %Nanjing, 211815, China. xinbingkong@126.com.
%}
%\and
%Yong-Xin Liu\thanks{School of Statistics and Data Science, Nanjing Audit %University, Nanjing, 211815,
%China. Corresponding and co-first author. liuyongxin@nau.edu.cn.}
%\and
% Long Yu\thanks{School of Statistics and Management, Shanghai University of %Finance and Economics, Shanghai, 200433, China. fduyulong@163.com. }
% \and
%Peng Zhao\thanks{School of Mathematics and Statistics, Jiangsu Normal %University, Xuzhou, 221116, China. zhaop@jsnu.edu.cn.
%}
%       }
%
%\date{}

%\maketitle

	\if1\blind
{
	\title{\bf Matrix Quantile Factor Model}
	\author{Xin-Bing Kong\\
		School of Statistics and Data Science \\
		Nanjing Audit University, Nanjing, 211815, China \\
		xinbingkong@126.com\\
		Yong-Xin Liu\\
		School of Statistics and Data Science \\Nanjing Audit University,
		 Nanjing, 211815,
		China\\
		 Corresponding and co-first author. liuyongxin@nau.edu.cn\\
		Long Yu \\
		School of Statistics and Management \\
		Shanghai University of Finance and Economics, Shanghai, 200433, China\\ fduyulong@163.com\\
		and\\
		Peng Zhao \\
		School of Mathematics and Statistics \\Jiangsu Normal University, Xuzhou, 221116, China\\ zhaop@jsnu.edu.cn}
	\maketitle
} \fi

\if0\blind
{
	\bigskip
	\bigskip
	\bigskip
	\begin{center}
		{\LARGE\bf  Matrix Quantile Factor Model}
	\end{center}
	\medskip
} \fi

\bigskip
\begin{abstract}
This paper introduces a matrix quantile factor model for matrix-valued data with low-rank structure. We estimate the row and column factor spaces via minimizing the empirical check loss function with orthogonal rotation constraints. We show that the estimates converge at rate $(\min\{p_1p_2,p_2T,p_1T\})^{-1/2}$ in the average Frobenius norm, where $p_1$, $p_2$ and $T$ are the row dimensionality, column dimensionality and length of the matrix sequence, respectively. This rate is faster than that of the quantile estimates via ``flattening" the matrix model into a large vector model. To derive the central limit theorem, we introduce a novel augmented Lagrangian function, which is equivalent to the original constrained empirical check loss minimization problem. Via the equivalence, we prove that the Hessian matrix of the augmented Lagrangian function is locally positive definite, resulting in a locally convex penalized loss function around the true factors and their loadings. This easily leads to a feasible second-order expansion of the score function and readily established central limit theorems of the smoothed estimates of the loadings. We provide three consistent criteria to determine the pair of row and column factor numbers. Extensive simulation studies and an empirical study justify our theory.
\end{abstract}

\noindent%
\vfill

%{ \emph{JEL Classification:}} C13, C31, C38

\newpage
\spacingset{1.9} % DON'T change the spacing!

%%%%%%%%%%%%%%%%%%%%%%%%%%%%%%%%%%%%%%%%%%%%%%%%%%%%%%%%%%%%%%%%%%%%%%%%%%%%%%

\section{Introduction}

% Problem and matrix data sequence

The present paper studies the matrix sequence data with a latent low-rank structure. Instead of modeling the mean functionals conditional on the latent factors as in recent works (\citet{wang2019factor}; \citet{yu2022projection}; \citet{Jing2021Community}), we model the conditional quantiles by an interactive effect of the row and column sections. The parameters, row and column factor loadings and factor matrices, are learnt by minimizing the empirical check loss under an identifiability constraint. For the first time, we derive the statistical accuracy of the estimated factor loadings and the factors. In the modeling side, this is essentially a work where the matrix factor structure meets the quantile feature representation.

On the matrix factor structure, we assume in the present paper that each matrix observation is driven by a much lower dimensional factor matrix, and that the two cross sections along the row and column dimensions interact with each other and thus generates the entries of the $t$-th data matrix $\bX_t=(X_{ijt})_{p_1\times p_2}$. For example, in recommending system, $\bX_t$ is a rating matrix of $p_1$ customers and $p_2$ commodities, and the scores in $\bX_t$ are high when the latent common consumption preferences of the customers match the latent common features of the $p_2$ goods. We focus on matrix sequence rather than large vectors appeared in standard statistics for two-fold reasons. First, many recent data sets in financial market, medical research, social networks and electronic business platform, are themselves well organized intrinsically in matrices. The matrix factor structure is empirically found in these data sets and works well in applications, c.f, \cite{Liu2023Simultaneous} for function magnetic resonance imaging data and \cite{Jing2022Community} for political blog network.
Second, modeling the matrix-value data with a low-rank structure, e.g. model (\ref{model}) below, makes the model parsimonious and statistical inference efficient once the structure is interpretably reasonable. A naive approach to analyze the data matrix $\bX_t$ is to ``flatten" it into a long vector $vec(\bX_t)$ by piling down column by column or row by row. After that, existing vector factor models in \cite{stock2002forecasting}, \cite{bai2002determining}, \cite{Trapani2018A}, \cite{Barigozzi2020Sequential}, \cite{fan2013large}, \cite{kong2017number}, \cite{kong2018systematic}, \cite{kong2019factor} and \cite{Chen2021quantile}, can be applied. However, the flattened vector factor modeling easily misses the interplay between the row and column sections, and has parameter complexity of order $O(p_1p_2+T)$ while the row-column interaction model (see (\ref{model}) below) of order $O(p_1+p_2+T)$. This is also where the efficiency gain of the present paper comes from compared with the vector quantile factor modeling.
For more motivation to study matrix or tensor sequence data, we refer to recent interesting works: \cite{wang2019factor}, \cite{chen2023testing}, \cite{Chang2023Modelling}, \cite{He2023Iterative},\cite{Yuan2023Two-way}, \cite{zhang2024tucker} and \cite{ZhangLiuGuoYuenWelsh2024JASA-modeling}.

On the quantile feature representation, mathematically, a $\tau$-quantile for a random variable $Y$ is $Q_{\tau}(Y)=\inf\{y; P(Y\leq y)\geq \tau\}$. With increasing complexity of data sets, how to understand the co-movement of the quantiles of large-dimensional random vectors evolving in time is of vital importance in theory and applications. To the best of our knowledge, \cite{Chen2021quantile} is the first paper that models the $\tau$ quantile of a large vector by a {\sc vector} factor structure. \cite{Ando2021Quantile} extended \cite{Chen2021quantile} to allow for observed covariates in modeling the panel quantiles. %\cite{Ando2019Pricing} applied the quantile factor structure to estimate the risk premium.
But, so far, no works are done to investigate the co-movement of the quanitles of a matrix sequence or even more generally tensor sequence. The more parsimonious interactive quantile factor representation, compared to the vector quantile factor model, is still not well understood in achieving higher statistical estimation precision. There is yet a challenge in establishing the second-order asymptotic theory for the estimated loadings, for example, the technique for the vector quantile factor model in the existing works can not be trivially extended to the matrix sequences. There is no guarantee that the Hessian matrix of the empirical check loss function penalized by the commonly used identifiability constraints is locally positive definite around the true parameters, making a second-order expansion of the penalized loss function difficult and hence the difficulty of the central limit theorems of the estimated row and column factor loadings.

In this paper, we estimate the row and column factor loadings and factors by minimizing the empirical check loss function under constraint. Our theory demonstrates that our estimates converge at rate $O_p((\min\{p_1T,p_2T,p_1p_2\})^{-1/2})$ in the sense of averaged Frobenius norm, if the quantile interactive mechanism is effective. Our theoretical rate is faster than $O_p((\min\{p_1p_2,T\})^{-1/2})$, the rate expected from the vector quantile factor analysis by vectorizing $\bX_t$, which is more pronounced when the sequence length $T$ is short. To the best of our knowledge, this is the first result on the estimation of the matrix quantile factor model and reveal of the interactive effect in reducing the estimation error. Our theory also shows that the convergence rates are reached without any moment constraints on the idiosyncratic errors, hence robust to the heavy tails of the heterogeneous idiosyncratic errors. To derive the central limit theorems of the smoothed versions of the loading estimates, we introduce an augmented Lagrangian function that not only takes the identification rotation constraints but also a cleverly constructed extra term into consideration. We have proved the reversibility of Hessian matrix for the smoothed quantile loss function with penalty locally around the true parameters, which is crucial to obtain the stochastic expansion of smoothed estimates and hence the central limit theorems. To solve the minimization of the check loss function under constraint, we present an iterative algorithm to find an approximate solution. Extensive simulation studies show that the numerical solutions are close enough to the true parameters, and demonstrate the robustness to the heavy tails. To determine the pair of the row and column factor numbers, we present three criteria, which are proved to be consistent and verified by simulations.

The present paper is organized as follows. Section \ref{MM} gives the matrix quantile factor model and the estimation method. Main results on estimating the cross-sectional factor spaces and set-up Assumptions are provided in Section \ref{MR}. Section \ref{No} presents three model selection criteria to determine the numbers of row and column factors. Section \ref{SEs} presents a smoothed version of the loading estimates and the central limit theorems. Section \ref{SE} conducts simulations and Section \ref{RD} presents an empirical data analysis. Section \ref{CD} concludes. The technical proofs are relegated to the supplementary material.

\section{Model and Methodology}\label{MM}

%The quantile was widely used in robust portfolio allocation, risk management, insurance regulation, quality evaluation, manufacturing monitoring, and so on. For example, in portfolio application, quantile-based scatter matrix can be used to construct robust portfolios.

We model the co-movement of the quantiles of all entries in each matrix by the following matrix quantile factor model.
\begin{eqnarray}\label{model}
\bX_t&=&Q_{\tau}(\bX_t|\bF_{t,\tau})+\bE_{t,\tau},\nonumber\\
Q_{\tau}(\bX_t|\bF_{t,\tau})&=&(Q_{\tau}(X_{ ijt}|\bF_{t,\tau}))_{p_1\times p_2}=\bR_{\tau}\bF_{t,\tau}\bC_{\tau}^{\prime},
\end{eqnarray}
where $\bR_{\tau}$, $\bC_{\tau}$ and $\bF_{t,\tau}$ are the $p_1\times k_{1,\tau}$ row factor loading matrix, $p_2\times k_{2,\tau}$ column factor loading matrix and $k_{1,\tau}\times k_{2,\tau}$ common factor matrix, respectively, and $\bE_{t,\tau}$ is an error matrix. Obviously, $Q_{\tau}(\bE_{t,\tau}|\bF_{t,\tau})=0$. The subscript $\tau$ emphasizes the dependence on $\tau$. That being said, the low-rank quantile structure is heterogeneous across different quantile levels, as seen in our real data analysis. Model (\ref{model}) demonstrates that the entries of $\bX_t$ depends on how close the rows of $\bR_{\tau}$ are to the rows of $\bC_{\tau}$, i.e, an interactive effect between the row and column sections of variables. We refer to $\bR_{\tau}\bF_{t,\tau}\bC_{\tau}^{\prime}$ and $\bE_{t,\tau}$ as the common and idiosyncratic components, respectively. Model (\ref{model}) includes the two-way quantile fixed effect model as a special example. In particular, setting $\bR_{\tau}=(\balpha_{p_1\times 1}(\tau), (\tilde{\bR}_{\tau})_{p_1\times (k_{1,\tau}-2)}(\tau), {\bf{1}}_{p_1\times 1})$, $\bF_{t,\tau}=\mbox{diag}\{1, (\tilde{\bF}_{t,\tau})_{(k_{1,\tau}-2)\times (k_{2,\tau}-2)}, 1\}$ and $\bC_{\tau}=({\bf{1}}_{p_2\times 1}, (\tilde{\bC}_{\tau})_{p_2\times (k_{2,\tau}-2)}, \bbeta_{p_2\times 1}(\tau))$,
$$
\bX_t=\balpha(\tau) {\bf{1}}_{1\times p_2}+{\bf{1}}_{p_1\times 1}\bbeta'(\tau)+\tilde{\bR}_{\tau}\tilde{\bF}_{t,\tau}\tilde{\bC}_{\tau}^{\prime}+\bE_{t,\tau},
$$
where $\balpha(\tau)$ and $\bbeta(\tau)$ represent the time-invariant quantile fixed effects along the row and column dimensions, respectively. They can be heterogeneous across the rows and/or columns.

While the vector factor model is conceptually a generative mechanism for a single cross-section of variables that are closely related in nature, the matrix factor model in (\ref{model}) is a two-way joint generative modeling in two totally different cross-section of variables. Though different in interpretations, model (\ref{model}) can be mathematically rewritten in the form of a vector factor model
\begin{equation}\label{vectorize}
\mbox{vec}(\bX_t)=(\bC_{\tau}\otimes\bR_{\tau})\mbox{vec}(\bF_{t,\tau})+\mbox{vec}(\bE_{t,\tau}),
\end{equation}
where $\mbox{vec}(\cdot)$ is the vectorization operator that stacks the columns of a matrix into a long vector and $\otimes$ stands for the Kronecker product operator. A general vector factor model for an observed vector $\bx_t$ is typically expressed as
\begin{equation}\label{vector}
(\bx_t)_{p\times 1}=\bL_{p\times k}(\bbf_t)_{k\times 1}+(\bepsilon_t)_{p\times 1},
\end{equation}
where $\bL$, $\bbf_t$ and $\bepsilon_t$ are the loading matrix, factor vector and idiosyncratic error vector, respectively. That is, (\ref{model}) can be mathematically regarded as a vector factor model with parameter restrictions $\bL=\bC_{\tau}\otimes \bR_{\tau}$, $p=p_1p_2$ and $k=k_1k_2$. When the Kronecker structure $\bC_{\tau}\otimes \bR_{\tau}$ is latent in the matrix sequence, a simple vectorization and vector principal component analysis would yield consistent estimate of the factor loading matrix $\bL$ (and hence $\bC_{\tau}\otimes \bR_{\tau}$) up to orthogonal transformation in the sense of averaged Frobenius norm. Expected from the vector quantile factor analysis in \cite{Chen2021quantile}, the consistent rate for estimating $\bL$ is $(\min\{p_1p_2,T\})^{-1/2}$. To recover the row and column factor spaces spanned by $\bR_{\tau}$ and $\bC_{\tau}$, a further nearest Kronecker decomposition has to be done, c.f., \cite{van2000kronecker}, but the resulting estimates of $\bR_{\tau}$ and $\bC_{\tau}$ depend on the estimation error for $\bL$. The other way around with vector quantile factor analysis is to minimize the empirical check loss function by restricting $\bL=\bC_{\tau}\otimes \bR_{\tau}$, but the number of restrictions is diverging which leads to complex computation. The matrix form (\ref{model}) gives a neat joint modeling of a two-way structure to start from.

Coming back to the general model (\ref{model}), the row factor loading matrix $\bR_{\tau}$, the column factor loading matrix $\bC_{\tau}$ and the factor matrix $\bF_{t, \tau}$ are not separately identifiable, though the common component itself is under some signal conditions. Indeed, there exists orthonormal square matrices $\bO_R$ and $\bO_C$, such that $\bR_{\tau}\bF_{t, \tau}\bC_{\tau}^{\prime}=\bR^*_{\tau}\bF^*_{t, \tau}\bC^{*\prime}_{\tau}$ where $\bR^*_{\tau}=\bR_{\tau}\bO_R$, $\bC^*_{\tau}=\bC_{\tau}\bO_C$ and $\bF_{t,\tau}^*=\bO_R^{\prime}\bF_{t,\tau}\bO_C^{\prime}$.
Without loss of generality, we assume throughout the paper that
\begin{equation}\label{id}
\frac{\bR^{\prime}_{\tau}\bR_{\tau}}{p_1}=\mathbb{I}_{k_1}, \ \frac{\bC^{\prime}_{\tau}\bC_{\tau}}{p_2}=\mathbb{I}_{k_2}, \ \frac{\sum^T_{t=1}\bF_{t,\tau}\bF_{t,\tau}^{\prime}}{T} \ \mbox{and} \ \frac{\sum^T_{t=1}\bF_{t,\tau}^{\prime}\bF_{t,\tau}}{T} \ \mbox{are diagonal matrices}.
\end{equation}

To estimate the parameters, we propose to minimize the empirical check loss function subject to (\ref{id})
\begin{equation}\label{check}
\min\mathbb{M}_{p_{1}p_{2}T}(\theta)=\frac{1}{p_1p_2T}\sum^{p_1}_{i=1}\sum^{p_2}_{j=1}\sum^T_{t=1}\rho_{\tau}(X_{ijt}-
\mathbf{r}_i^{\prime}\bF_t\mathbf{c}_j)  \ \mbox{subject to (\ref{id})},
\end{equation}
with respect to $\theta=\{\mathbf{r}_1,..., \mathbf{r}_{p_1};\mathbf{c}_1,...,\mathbf{c}_{p_2};\bF_1,...,\bF_T\}$, where $\rho_{\tau}(u)=(\tau-I\{u\leq 0\})u$, and $\mathbf{r}_i^{\prime}$ and $\mathbf{c}_j^{\prime}$ are the $i$-th row of $\bR_{\tau}$ and $j$-th row of $\bC_{\tau}$, respectively. Our estimates, denoted by $\hat{\bR}_{\tau}$, $\hat{\bF}_{t,\tau}$ and $\hat{\bC}_{\tau}$, are simply the minimizers of the above empirical check loss function assuming that $k_{1, \tau}$ and $k_{2, \tau}$ are known numbers of factors a priori. Later, we will give consistent estimates of $k_{1, \tau}$ and $k_{2, \tau}$ by three methods. Notice that the empirical check loss function is not a convex function jointly in $\bR_{\tau}$, $\bF_{t, \tau}$
and $\bC_{\tau}$, but it is a marginally convex function when the other two are fixed. Hence, we propose to optimize it via an iterative algorithm; see Algorithm \ref{alg1} below.

\begin{algorithm}[H]
	\caption{Iterative algorithm for the row and column factor loading matrices and the factor matrix}\label{alg1}
	{\bf Input:} Data matrices $\{\bX_t\}_{t\le T}$, the pair of row and column factor numbers $k_1$ and $k_2$\\
	{\bf Output:} Factor loading matrices and factor matrix

Step I: set $h=0$ and give initial values of $\{\hat{\bF}_t(0)\}_{t=1}^T$ and $\hat{\bC}(0)$ satisfying (\ref{id});
		
Step II: given $\{\hat{\bF}_t(h)\}_{t=1}^T$ and $\hat{\bC}(h)$, minimize $\mathbb{M}_{p_1p_2T}(\theta)$ with respect to $\bR$ and obtain a normalized $\hat{\bR}(h+1)$ so that (\ref{id}) is fulfilled;
		
Step III: given $\hat \bR(h+1)$ and $\hat{\bC}(h)$, minimize $\mathbb{M}_{p_1p_2T}(\theta)$ with respect to $\bF_1,\ldots,\bF_T$ and obtain $\{\hat{\bF}_t(h+1)\}_{t=1}^T$;

Step IV: given $\hat \bR(h+1)$ and $\{\hat{\bF}_t(h+1)\}_{t=1}^T$, minimize $\mathbb{M}_{p_1p_2T}(\theta)$ with respect to $\bC$ and obtain a normalized $\hat\bC(h+1)$ so that (\ref{id}) is fulfilled;

Step V: set $h=h+1$ and repeat Steps II to IV until convergence or up to $h=m$.

\end{algorithm}

Although $\mathbb{M}_{p_1p_2T}(\theta)$ is not a joint convex function, it is convex in each iteration in one component of $(\hat{\bR}(h), \hat{\bC}(h), \hat{\bF}_t(h))$ with the other two given. Motivated by \cite{Ge2017No}, we set the initial values in Algorithm \ref{alg1} by random initialization. Our simulation shows that the algorithm converges fast and leads to accurate estimation.

\section{Estimation of the Factor Spaces}\label{MR}

In this section, we present a main result on the estimation accuracy of the estimated row and column factor loading matrices. Before stating the theorem, we give some technical assumptions. Without confusion, we suppress the dependence on $\tau$ of the notation $k_{1,\tau}$ and $k_{2,\tau}$, and write them simply as $k_1$ and $k_2$.

\begin{assumption}\label{ass1}
Let $\mathcal{A}\subset \mathbb{R}^{k_{1}}$, $\mathcal{F}\subset \mathbb{R}^{k_{1}\times k_{2}}$,
$\mathcal{B}\subset \mathbb{R}^{k_{2}}$ and define
\begin{equation}
\Theta^{k_{1}k_{2}}=\{ \mathbf{r}_{i}\in \mathcal{A}, \bF_{t,\tau}\in \mathcal{F}, \mathbf{c}_{j}\in \mathcal{B} \ \mbox{satisfying} \ (\ref{id})\}.
\end{equation}
\begin{enumerate}
\item $\mathcal{A}$, $\mathcal{F}$, $\mathcal{B}$ are compact sets
and the true parameter $\theta_{0}\in \Theta^{k_{1}k_{2}}$. The true factor matrix $\bF^0_{t,\tau}$ satisfies
\begin{equation}\label{F1}
\frac{1}{T}\sum_{t=1}^{T}\bF_{t,\tau}^0\bF_{t,\tau}^{0\prime}
=\mbox{diag}(\sigma_{T1},\cdots,\sigma_{Tk_{1}})
\end{equation}
with $\sigma_{T1}\geq \cdots\geq \sigma_{Tk_{1}}$
and $\sigma_{Tj}\rightarrow \sigma_{j}$ as $T\rightarrow\infty$
for $j=1,\ldots,k_{1}$ with $\infty>\sigma_{1}>\cdots>\sigma_{k_{1}}>0$.
\begin{equation}\label{F2}
\frac{1}{T}\sum_{t=1}^{T}\bF_{t,\tau}^{0\prime}\bF_{t,\tau}^0
=\mbox{diag}(\tilde{\sigma}_{T1},\cdots,\tilde{\sigma}_{Tk_{2}})
\end{equation}
with $\tilde{\sigma}_{T1}\geq \cdots\geq \tilde{\sigma}_{Tk_{2}}$
and $\tilde{\sigma}_{Tj}\rightarrow \tilde{\sigma}_{j}$ as $T\rightarrow\infty$
for $j=1,\ldots,k_{2}$ with $\infty>\tilde{\sigma}_{1}>\cdots>\tilde{\sigma}_{k_{2}}>0$.

\item The conditional density function of the idiosyncratic error variable $\varepsilon_{ijt}$ given $\{\bF_{t,\tau}^0\}$, denoted as $\text{f}_{ijt}$, is continuous, and satisfies that: for any compact set $\mathcal{I} \subset R$ and any $x\in \mathcal{I}$, there exists a positive constant $\underline{\text{f}}>0$ such that $\text{f}_{ijt}(x)\geq \underline{\text{f}}$ for all $i,j,t$.

\item Given $\{\bF_{t,\tau}^0, 1\leq t \leq T \}$, $\{\varepsilon_{ijt}, 1\leq i \leq p_{1}, 1\leq j \leq p_{2}, 1\leq t \leq T\}$ are independent across $i,j$ and $t$.
\end{enumerate}
\end{assumption}

Assumption \ref{ass1}-1 is standard in the literature, e.g., the compactness of the parameters were assumed in \cite{Chen2021quantile}, and the existence of the limits in (\ref{F1}) and (\ref{F2}) is guaranteed by the law of large numbers under various weak-correlation conditions. Assumption \ref{ass1}-2 assumed the existence of density functions which are uniformly bounded from below in compact sets, see also similar conditions in \cite{Chen2021quantile}. Assumption \ref{ass1}-3 restricts that the idiosyncratic errors are conditionally independent but maybe dependent unconditionally, see the same condition in \cite{Chen2021quantile}. Even if (\ref{id}) is satisfied, the columns of  loading matrices $\bR_{\tau}$ and $\bC_{\tau}$ are identifiable only up to a positive or negative sign. We henceforth make a convention that the first nonzero entry of each column of $\bR_{\tau}$ and $\bC_{\tau}$ is positive.

\begin{theorem}\label{th1}
Under Assumption \ref{ass1}, as $p_1, p_2, T\rightarrow \infty$,
$$
\frac{\|\hat{\bR}_{\tau}-\bR_{0,\tau}\|_F}{\sqrt{p_{1}}}+
\frac{\|\hat{\bF}_{\tau}-\bF_{0,\tau}\|_F}{\sqrt{T}}+
\frac{\|\hat{\bC}_{\tau}-\bC_{0,\tau}\|_F}{\sqrt{p_{2}}}=O_{p}(L_{p_{1}p_{2}T}^{-1}),
$$
where $L_{p_1p_2T}=(\min\{p_1p_2,p_2T,p_1T\})^{1/2}$, and $\bR_{0,\tau}$, $\bF_{0,\tau}$ and $\bC_{0,\tau}$ are, respectively, the true row factor loading matrix, the factor matrix and the column factor loading matrix.
\end{theorem}

Theorem \ref{th1} demonstrates that the plug-in estimate $\hat{\bC}_{\tau}\otimes\hat{\bR}_{\tau}$ has convergence rate:
\begin{eqnarray}\label{plug-in}
&&\|\hat{\bC}_{\tau}\otimes\hat{\bR}_{\tau}-\bC_{0,\tau}\otimes \bR_{0, \tau}\|_F/(p_1p_2)^{1/2}=O_p(L_{p_1p_2T}^{-1}).
\end{eqnarray}
Expected from \cite{Chen2021quantile}, the convergence rate of estimating the loading space spanned by $\bC_{0,\tau}\otimes\bR_{0,\tau}$ under the framework of the vector quantile factor model (\ref{vector}) is $O_p((\min\{p_1p_2,T\})^{-1/2})$ by piling down the columns of each observed matrix into a long vector. A simple comparison shows that the latter rate is no faster than ours, and in particular, when $p_1p_2$ dominates $T$, ours is strictly faster than the rate by vectorizing the matrix. This is intuitively interpretable because the structure restriction of $\bL$ in (\ref{vector}) is not observed in the vector quantile analysis.

\section{Model Selection Criteria}\label{No}

\subsection{Rank Minimization}
We propose three different methods to select the numbers of factors. The first selects the numbers of factors by rank minimization (RM), the second uses the information criterion (IC), while the third implements the eigenvalue ratio thresholding approach (ER). As before, the dependence on $\tau$ in all mathematical notations are suppressed for simplicity.

Let $K_{1}$ and $K_{2}$ be two positive integers larger than $k_{1}$ and $k_{2}$, respectively.
Let $\mathcal{A}^{K_{1}}$ be compact subset of $\mathbb{R}^{K_{1}}$, $\mathcal{F}^{K_{1}\times K_{2}}$ be compact subset of $\mathbb{R}^{K_{1}\times K_{2}}$ and $\mathcal{B}^{K_{2}}$ be compact subset of $\mathbb{R}^{K_{2}}$. Assume that
\begin{equation*}
\begin{aligned}
\left(
        \begin{array}{cc}
          \bF_{0t} &   \bf{0}_{k_{1}\times (K_{2}-k_{2})} \\
          \bf{0}_{(K_{1}-k_{1}) \times k_{2}} & \bf{0}_{(K_{1}-k_{1}) \times (K_{2}-k_{2})}
        \end{array}
      \right)\in \mathcal{F}^{K_{1}\times K_{2}}
\end{aligned}
\end{equation*}
for all $t$, $(\mathbf{r}_{0i}, \bf{0})\in \mathcal{A}^{K_1}$ and $(\mathbf{c}_{0j}, \bf{0})\in\mathcal{B}^{K_2}$. Let $\mathbf{r}_{i}^{K_{1}} \in \mathbb{R}^{K_{1}}$, $\bF_{t}^{K_{1}\times K_{2}} \in \mathbb{R}^{K_{1}\times K_{2}}$, $\mathbf{c}_{j}^{K_{2}} \in \mathbb{R}^{K_{2}}$ for all $i, t, j$ and
write
\begin{eqnarray*}
\mathbf{\theta}^{K_{1}K_{2}}&=&(({\mathbf{r}_{1}^{K_{1}}})^{\prime},\cdots,({\mathbf{r}_{p_{1}}^{K_{1}}})^{\prime},
\bF_{1}^{K_{1}\times K_{2}} ,\cdots,\bF_{T}^{K_{1}\times K_{2}} ,({\mathbf{c}_{1}^{K_{2}}})^{\prime},\cdots,({\mathbf{c}_{p_{2}}^{K_{2}}})^{\prime})',\\
\bR^{K_{1}}&=&(\mathbf{r}_{1}^{K_{1}},\cdots,\mathbf{r}_{p_{1}}^{K_{1}})^{\prime}, \bF^{K_{1}\times K_{2}}=(\bF_{1}^{K_{1}\times K_{2}},\cdots,\bF_{T}^{K_{1}\times K_{2}}), \bC^{K_{2}}=(\mathbf{c}_{1}^{K_{2}},\cdots,\mathbf{c}_{p_{2}}^{K_{2}})^{\prime}.
\end{eqnarray*}

Consider the following normalization,
\begin{eqnarray}\label{6}
&&\frac{1}{p_{1}}(\bR^{K_{1}})'\bR^{K_{1}}=\mathbb{I}_{K_{1}}, \frac{1}{p_{2}}(\bC^{K_{2}})'\bC^{K_{2}}=\mathbb{I}_{K_{2}};\nonumber\\
&&\frac{1}{T}\sum_{t=1}^{T}\bF_{t}^{K_{1}\times K_{2}}(\bF_{t}^{K_{1}\times K_{2}})' \ \mbox{and} \ \frac{1}{T}\sum_{t=1}^{T}(\bF_{t}^{K_{1}\times K_{2}})'\bF_{t}^{K_{1}\times K_{2}} \
\text{diagonal.}
\end{eqnarray}
Define
\begin{equation*}
\begin{aligned}
\mathbf{\Theta}^{K_{1}K_{2}}=\{\theta^{K_{1}K_{2}}:& \ \mathbf{r}_{i}^{K_{1}}\in \mathcal{A}^{K_{1}}, \bF_{t}^{K_{1}\times K_{2}}\in \mathcal{F}^{K_{1}\times K_{2}}, \mathbf{c}_{j}^{K_{2}}\in \mathcal{B}^{K_{2}} \ \text{for all} \ i,t,j; \\
& \mathbf{r}_{i}^{K_{1}},\bF_{t}^{K_{1}\times K_{2}}\ \mbox{and} \ \mathbf{c}_{j}^{K_{2}}\text{ satisfy (\ref{6})}\},
\end{aligned}
\end{equation*}
and
\begin{equation*}
%\left\{
\begin{aligned}
\hat{\mathbf{\theta}}^{K_{1}K_{2}}&=( (\widehat{\mathbf{r}}_{1}^{K_{1}})',\cdots,(\widehat{\mathbf{r}}_{p_{1}}^{K_{1}})',
\widehat{\bF}_{1}^{K_{1}\times K_{2}} ,\cdots,\widehat{\bF}_{T}^{K_{1}\times K_{2}} ,(\widehat{\mathbf{c}}_{1}^{K_{2}})',\cdots,(\widehat{\mathbf{c}}_{p_{2}}^{K_{2}})')'\\
&=\arg\min_{\mathbf{\theta}^{K_{1}K_{2}}\in\mathbf{\Theta}^{K_{1}K_{2}}}
\frac{1}{p_{1}p_{2}T}\sum_{i=1}^{p_{1}}\sum_{j=1}^{p_{2}}\sum_{t=1}^{T}
\rho_{\tau}(X_{ijt}-(\mathbf{r}_{i}^{K_{1}})'\bF_{t}^{K_{1}\times K_{2}}\mathbf{c}_{j}^{K_{2}}).
\end{aligned}
%\right.
\end{equation*}
Moreover, write $\widehat{\bF}^{K_{1}\times K_{2}}=(\widehat{\bF}^{K_{1}\times K_{2}}_{1},\cdots,\widehat{\bF}^{K_{1}\times K_{2}}_{T})$ and
\begin{equation*}
%\left\{
\begin{aligned}
\frac{1}{T}\sum_{t=1}^{T}\widehat{\bF}^{K_{1}\times K_{2}}_{t}(\widehat{\bF}^{K_{1}\times K_{2}}_{t})'
&=\text{diag}(\widehat{\sigma}^{K_{1}}_{T,1},\cdots,\widehat{\sigma}^{K_{1}}_{T,K_{1}}),\\
\frac{1}{T}\sum_{t=1}^{T}(\widehat{\bF}^{K_{1}\times K_{2}}_{t})'\widehat{\bF}^{K_{1}\times K_{2}}_{t}
&=\text{diag}(\widehat{\sigma}^{K_{2}}_{T,1},\cdots,\widehat{\sigma}^{K_{2}}_{T,K_{2}}).
\end{aligned}
%\right.
\end{equation*}

The rank minimization estimator of the numbers of factors, $k_{1}$ and $k_{2}$, are defined as
\begin{equation*}
\begin{aligned}
\widehat{k}_{1}^{r}&=\sum_{j=1}^{K_{1}}\textbf{1}\{\widehat{\sigma}^{K_{1}}_{T,j}>C_{p_{1}p_{2}T} \}, \
\widehat{k}_{2}^{r}&=\sum_{j=1}^{K_{2}}\textbf{1}\{\widehat{\sigma}^{K_{2}}_{T,j}>C_{p_{1}p_{2}T} \},
\end{aligned}
\end{equation*}
where $C_{p_{1}p_{2}T}$ is a sequence that goes to 0 as $p_{1},p_{2},T\rightarrow \infty$. That being said, $\widehat{k}_{1}^r$ and $\widehat{k}_{2}^r$ are, respectively, the
numbers of the diagonal elements of
$$
\sum_{t=1}^{T}\widehat{\bF}^{K_{1}\times K_{2}}_{t}(\widehat{\bF}^{K_{1}\times K_{2}}_{t})'/T,\quad \sum_{t=1}^{T}(\widehat{\bF}^{K_{1}\times K_{2}}_{t})'\widehat{\bF}^{K_{1}\times K_{2}}_{t}/T
$$
that are larger than the threshold $C_{p_{1}p_{2}T}$.

\begin{theorem}\label{th3.1}
Under Assumption \ref{ass1},
$
P\left(\widehat{k}_{1}^r=k_{1}, \widehat{k}_{2}^r=k_{2}\right)\rightarrow 1
$
as $p_{1},p_{2},T\rightarrow \infty$ if
$K_{1}>k_{1}$, $K_{2}>k_{2}$, $C_{p_{1}p_{2}T}\rightarrow 0$, and $C_{p_{1}p_{2}T}L_{p_{1}p_{2}T}^{2}\rightarrow \infty$.
\end{theorem}

\subsection{Information Criterion}

The second estimator of $(k_{1}, k_{2})$ is similar to the IC-based estimator of \cite{bai2002determining}, but is adaptive to the matrix observation and the check loss function. For $(l_1, l_2)\in \mathcal{P}=\{0,..., K_1\}\times \{0,..., K_2\}$, we search the minimizer of a penalized empirical check loss function.

The IC-based estimator of $(k_{1}, k_{2})$ is defined as
\begin{equation*}
(\widehat{k}_{1}^{IC},\widehat{k}_{2}^{IC})=
\arg\min_{(l_1,l_2)\in\mathcal{P}}\left(\mathbb{M}_{p_{1}p_{2}T}(\widehat{\mathbf{\theta}}^{l_{1}l_{2}})+(l_{1}+l_{2}) C_{p_{1}p_{2}T}\right),
\end{equation*}
where $\hat{\mathbf{\theta}}^{l_1l_2}$ is similarly defined as $\hat{\mathbf{\theta}}^{K_1K_2}$ except for  replacing $(K_1, K_2)$ by $(l_1, l_2)$, pretending that there are $l_1$ row factor and $l_2$ column factors.

\begin{theorem}\label{th3.2}
Suppose Assumption \ref{ass1} holds, and assume that for any compact set $\mathcal{I} \in \mathbb{R}$
and any $u\in \mathcal{I}$, there exists $\overline{f}$ such that $\text{f}_{ijt}(u)\leq \overline{f}$ for all $i, j, t$. Then
$P\left(\widehat{k}_{1}^{IC}=k_{1},\widehat{k}_{2}^{IC}=k_{2}\right)$ $\rightarrow1$, as $p_{1},p_{2},T\rightarrow \infty$ if $C_{p_{1}p_{2}T}\rightarrow 0$ and $C_{p_{1}p_{2}T}L_{p_{1}p_{2}T}^{2}\rightarrow \infty$.
\end{theorem}

\subsection{Eigenvalue Ratio Thresholding}

Due to the assumption of $\bF_{t}^{K_{1}\times K_{2}}$ in section 4.1, we expect $(\hat\sigma_{T,k_1+1}^{K_1},\ldots,\hat\sigma_{T,K_1}^{K_1})$ and $(\hat\sigma_{T,k_2+1}^{K_2},\ldots,\hat\sigma_{T,K_2}^{K_2})$ to be redundant and negligible. Therefore, motivated by the eigenvalue ratio approach in \cite{Ahn2013eigenvalue},
a direct estimator for $(k_1, k_2)$ is given by
$$
\hat k_1^{ER}=\arg\max_{1\le k\le K_1-1}\frac{\hat\sigma_{T,k}^{K_1}}{\hat\sigma_{T,k+1}^{K_1}+c_0L_{p_1p_2T}^{-2}}, \
\hat k_2^{ER}=\arg\max_{1\le k\le K_2-1}\frac{\hat\sigma_{T,k}^{K_2}}{\hat\sigma_{T,k+1}^{K_2}+c_0L_{p_1p_2T}^{-2}},
$$
where $c_0$ is a small positive constant so that the denominator is always larger than 0. In our simulation studies and real data analysis, we set $c_0=10^{-4}$.

\begin{theorem}\label{ER}
	Under Assumption \ref{ass1}, as $p_1, p_2, T\rightarrow \infty$ and $c_0\rightarrow 0$,
$$
P\left(\hat k_1^{ER}=k_1,  \hat k_2^{ER}=k_2\right)\rightarrow 1.
$$
\end{theorem}

\section{Smoothed Estimates}\label{SEs}
The non-smoothness of the check loss function and the incidental-parameter problem make it difficult to derive the asymptotic distribution of the estimators $\widehat{\theta}$. As in the asymptotic analysis of quantile regression, one way to overcome these difficulties is to expand the expected score function and obtain a stochastic expansion for $\widehat{\mathbf{r}}_{i}-\mathbf{r}_{0i}$.

We proceed by defining a new estimator of $\mathbf{\theta}_{0}$, denoted as $\widetilde{\mathbf{\theta}}$
which relies on the following smoothed quantile optimization:
\begin{equation}\label{smooth}
\widetilde{\mathbf{\theta}}=(\widetilde{\mathbf{r}}_{1}',\ldots,\widetilde{\mathbf{r}}_{p_{1}}',
\widetilde{\bF}_{1},\ldots,\widetilde{\bF}_{T},\widetilde{\mathbf{c}}_{1}',\ldots,\widetilde{\mathbf{c}}_{p_{2}}')'
=\arg\min_{\mathbf{\theta} \in \mathbf{\Theta}^{k_{1}k_{2}}} \mathbb{S}_{p_{1}p_{2}T}(\mathbf{\theta}) \ \mbox{subject to (\ref{id})},
\end{equation}
where
$$
\mathbb{S}_{p_{1}p_{2}T}(\theta)=\frac{1}{p_{1}p_{2}T}\sum_{i=1}^{p_{1}}\sum_{j=1}^{p_{2}}\sum_{t=1}^{T}
\Big[\tau-K\Big(\frac{X_{ijt}-\mathbf{r}_{i}'\bF_{t}\mathbf{c}_{j}}{h}\Big)\Big](X_{ijt}-\mathbf{r}_{i}'\mathbf{F}_{t}\mathbf{c}_{j}),
$$
such that $K(z)=1-\int_{-1}^{z}k(z)dz$, $k(z)$ is a continuous kernel function with support $[-1,1]$ and $h$ is a bandwidth parameter that goes to 0 as $p_{1},p_{2}$ and $T$ grow.

 To derive the central limit theorem, we instead introduce an augmented Lagrangian function which is equivalent to (\ref{smooth}), that is
\begin{equation}\label{Lagrangian}
\widetilde{\mathbf{\theta}}=
\arg\min_{\mathbf{\theta} \in \mathbf{\Theta}^{k_{1}k_{2}}} \mathbb{S}_{p_{1}p_{2}T}(\mathbf{\theta}) + \mathbb{P}_{p_{1}p_{2}T}(\mathbf{\theta}),
\end{equation}
where
\begin{eqnarray*}
&&\mathbb{P}_{p_{1}p_{2}T}(\mathbf{\theta})=\mathbb{P}_{1}(\mathbf{\theta})+\mathbb{P}_{2 }(\mathbf{\theta})+\mathbb{P}_{3}(\mathbf{\theta})\\
&=&b_1\Bigg[\frac{1}{2p_1}\sum_{p=1}^{k_1}\sum_{q>p}^{k_1}\left(\sum_{i=1}^{p_1}r_{ip}r_{iq}\right)^2 + \frac{1}{8p_1}\sum_{k=1}^{k_1}\left(\sum_{i=1}^{p_1}r_{ik}^2-p_1\right)^2+\frac{1}{2T}\sum_{p=1}^{k_1}\sum_{q>p}^{k_1}
\left(\sum_{t=1}^T\mathbf{F}_{tp\cdot}\mathbf{F}_{tq\cdot}^{\prime}\right)^2\Bigg]\\
&+&b_2\Bigg[\frac{1}{2p_2}\sum_{p=1}^{k_2}\sum_{q>p}^{k_2}\left(\sum_{j=1}^{p_1}c_{jp}c_{jq}\right)^2 + \frac{1}{8p_2}\sum_{k=1}^{k_2}\left(\sum_{j=1}^{p_2}c_{jk}^2-p_2\right)^2+ \frac{1}{2T}\sum_{p=1}^{k_2}\sum_{q>p}^{k_2}\left(\sum_{t=1}^T\mathbf{F}_{t\cdot p}^{\prime}\mathbf{F}_{t\cdot q}\right)^2\Bigg]\\
&+&b_3\sum_{t=1}^T\Bigg[\frac{1}{2p_1}\sum_{p=1}^{k_1}\sum_{q=1}^{k_2}\left(\sum_{i=1}^{p_1}\left(\frac{r_{ip}^2-p_{1}}{2}
f_{t,pq} + \sum_{k\neq p}^{k_1}r_{ip}r_{ik}f_{t,kq}\right)\right)^2\Bigg],
\end{eqnarray*}
where $b_1$, $b_2$ and $b_3$ are positive Lagrangian multipliers, $\mathbf{F}_{tp.}$ is the $p$-th row of $\mathbf{F}_{t}$, $\mathbf{F}_{t.q}$ is the $q$-th column of $\bF_{t}$ and $f_{t,pq}$ is the element of the $p$-th row and $q$-th column of $\bF_{t}$.

The equivalence between (\ref{Lagrangian}) and (\ref{smooth}) stems from the nonnegative property of $\mathbb{P}_{p_{1}p_{2}T}(\mathbf{\theta})$. Theoretically the minima of (\ref{smooth}) are achieved if and only if $\mathbb{P}_{p_{1}p_{2}T}(\mathbf{\theta})=0$. The first two terms $\mathbb{P}_{1}(\mathbf{\theta})$ and $\mathbb{P}_{2}(\mathbf{\theta})$ are associated with the typical four rotation constraints in (\ref{id}) under which the row and column factor loadings and the factor matrices are uniquely identified up to signs. The additional augmented term $\mathbb{P}_{3}(\mathbf{\theta})$, however, is carefully designed to ensure positive definiteness of the Hessian matrix of the penalized loss function, making the augmented Lagrangian function (\ref{Lagrangian}) convex locally in $R$, $C$ and $F_t$'s around their true values. This renders a feasible second-order expansion of the estimation errors which makes the central limit theorems of the estimates easily derived.
Since in the matrix factor model (\ref{model}), there are two cross-sections, row and column, only making use of the rotation identification constraint, i.e. $\mathbb{P}_{1}(\mathbf{\theta})+\mathbb{P}_{2}(\mathbf{\theta})$, as traditionally done in the quantile {\sc vector} factor model, is difficult to prove the local convexity of the penalized loss function. The proposed augmented Lagrangian loss function (\ref{Lagrangian}) skillfully solved this problem, see the Appendix for the technical details.

Before stating the central limit theorem, we define, for all $i, j, t$.
  \begin{align*}
  \Phi_{i}&=\lim_{T\rightarrow \infty}\lim_{p_{2}\rightarrow \infty}\frac{1}{Tp_{2}}
\sum_{t=1}^{T}\sum_{j=1}^{p_{2}}f_{ijt}(0)\bF_{0t}\mathbf{c}_{0j}\mathbf{c}_{0j}'\bF_{0t}',\\
\Psi_{t}&=\lim_{p_{1}\rightarrow \infty}\lim_{p_{2}\rightarrow \infty}\frac{1}{p_{1}p_{2}}
\sum_{i=1}^{p_{1}}\sum_{j=1}^{p_{2}}f_{ijt}(0)(\mathbf{c}_{0j}\otimes \mathbf{r}_{0i})(\mathbf{c}'_{0j}\otimes \mathbf{r}'_{0i}),\\
\varphi_{j}&=\lim_{T\rightarrow \infty}\lim_{p_{1}\rightarrow\infty}
\frac{1}{Tp_{1}}\sum_{t=1}^{T}\sum_{i=1}^{p_{1}}f_{ijt}(0)
\bF_{0t}'\mathbf{r}_{0i}\mathbf{r}_{0i}'\bF_{0t}.
  \end{align*}

\begin{assumption}\label{ass2}
Let $m\geq 8$ be a positive integer,
\begin{enumerate}
\item $\Phi_{i}>0$, $\Psi_{t}>0$, $\varphi_{j}>0$ for all $i, t, j$.

\item $\mathbf{r}_{0i}$ is an interior point of $\mathcal{A}$, $\mathbf{c}_{0j}$ is an interior point of $\mathcal{B}$ and $\bF_{0t}$ is an interior point of $\mathcal{F}$ for all $i, j, t$.

\item $k(z)$ is symmetric and twice differentiable. For $s=1,\ldots,m-1$, $\int_{-1}^{1}k(z)dz=1$,  $\int_{-1}^{1}z^{s}k(z)dz=0$, and $\int_{-1}^{1}z^{m}k(z)dz\neq 0$.

\item $f_{ijt}$ is $m+2$ times continuously differentiable. Let $f_{ijt}^{(s)}(u)=(\partial/\partial u)^{s}f_{ijt}(u)$ for $s=1,\ldots,m+2$.
For any compact set $\mathcal{I}\subset \mathbb{R}$ and any $u \in \mathcal{I}$, there exist $-\infty<\underline{l}<\bar{l}$ (depending on $\mathcal{I}$) such that $\underline{l}<f_{ijt}^{(s)}(u)<\bar{l}$, $\underline{f}\leq f_{ijt}(u)\leq\bar{l}$
for $s=1,\ldots,m+2$ and for all $i, j, t$.

\item as $p_{1},p_{2},T\rightarrow \infty$, $T\propto p_{1}$, $T\propto p_{2}$, $p_{1}\propto p_{2}$, $h\propto T^{-2c}$ and $m^{-1}<c<1/6$.
\end{enumerate}
\end{assumption}

The above conditions are standard in smoothed quantile optimization, with the exception of Assumption \ref{ass2}-5. Note that, as in
Galvao and Kato (2016), we require $k(z)$ to be a higher-order kernel function to control
the higher-order terms in the stochastic expansions of the estimators. However, Galvao
and Kato (2016) assumed that $m^{-1}<c<1/3$, while we need $m^{-1}<c<1/6$
. This arises from the fact that the incidental parameters, $\mathbf{r}_{0i}$, $\bF_{0t}$ and
$\mathbf{c}_{0j}$, in quantile factor models enter the model interactively, but no interactive fixed effects appear in the panel quantile models considered by these authors.

\begin{theorem}\label{CLT}
 Under Assumptions \ref{ass1} and \ref{ass2},
 \begin{eqnarray*}
 	(Tp_{2})^{1/2}(\widetilde{\mathbf{r}}_{i}-\mathbf{r}_{0i})\rightarrow \mathcal{N}(0,\tau(1-\tau)\Phi_{i}^{-1}\Sigma_{1}\Phi_{i}^{-1}), \\
 	(Tp_{1})^{1/2}(\widetilde{\mathbf{c}}_{j}-\mathbf{c}_{0j})\rightarrow \mathcal{N}(0,\tau(1-\tau)\varphi_{j}^{-1}\Sigma_{2}\varphi_{j}^{-1}),
 \end{eqnarray*}
where $\Sigma_{1}=\sum_{t=1}^{T}\bF_{t,\tau}^0\bF_{t,\tau}^{0\prime}/T$,
$\Sigma_{2}=\sum_{t=1}^{T}\bF_{t,\tau}^{0\prime}\bF_{t,\tau}^0/T$.
\end{theorem}

\begin{remark}
Similar to the proof of Theorem 1, it holds that
\begin{equation*}
\begin{aligned}
\frac{\|\widetilde{\bR}_{\tau}-\bR_{0,\tau}\|_F}{p_1^{1/2}}=O_{p}&(L_{p_{1}p_{2}T}^{-1})+O_{p}(h^{m/2}), \quad
\frac{\|\widetilde{\bF}_{\tau}-\bF_{0,\tau}\|_F}{T^{1/2}}=O_{p}(L_{p_{1}p_{2}T}^{-1})+O_{p}(h^{m/2}), \\
&\frac{\|\widetilde{\bC}_{\tau}-\bC_{0,\tau}\|_F}{p_2^{1/2}}=O_{p}(L_{p_{1}p_{2}T}^{-1})+O_{p}(h^{m/2}),
\end{aligned}
\end{equation*}
where the extra $O_{p}(h^{m/2})$ term is due to the approximation bias of the smoothed check
function. However, Assumption \ref{ass2}-5 implies that $1/L_{p_{1}p_{2}T}\gg h^{m/2}$, and then it follows that
average convergence rates of $\widetilde{\bR}_{\tau}$, $\widetilde{\bF}_{\tau}$ and $\widetilde{\bC}_{\tau}$ are all $L_{p_{1}p_{2}T}$.
\end{remark}

 \section{Simulation Studies}\label{SE}

\subsection{Data generating process}

We generate data from the following matrix series,
\begin{equation}\label{data generating}
\bX_t=\bR\bF_t\bC^\prime+\theta^* g_t\bE_t,
\end{equation}
where $\bR$ and $\bC$ are $p_1\times k_1$ and $p_2\times k_2$ matrices, respectively. We set $k_1=2$ and $k_2=3$. The factor process follows an autoregressive model such that $\bF_t=0.2\bF_{t-1}+\Xi_t$.
 $g_t$ is a scalar random variable satisfying $g_t=0.2g_{t-1}+\epsilon_t$. The entries in $\bR$, $\bC$, $\{\Xi_t\}$ and $\{\epsilon_t\}$ are all generated from i.i.d. $\mathcal{N}(0,1)$. The entries of $\{\bE_t\}$ are i.i.d. from $\mathcal{N}(0,1)$, or $t$ distributions with degree of freedom being 3 or 1, covering both light-tailed and heavy-tailed distributions. $\theta^*$ is a parameter controlling the signal-to-noise ratio.

 To ensure the identification condition (\ref{id}), a normalization step should be applied to the loading and factor score matrices. For instance, when $\tau=0.5$, do singular-value decomposition to $\bR$ and $\bC$ as
 \[
 \bR=\bU_R\bD_R{\mbox{\boldmath $V$}}_R=\bU_R\bQ_R,\quad  \bC=\bU_C\bD_C{\mbox{\boldmath $V$}}_C=\bU_C\bQ_C.
 \]
 Further define
 \[
\tilde\bSigma_1=\frac{1}{Tp_1p_2}\sum_{t=1}^T\bQ_R\bF_t\bC^\prime\bC\bF_t^\prime\bQ_R^\prime,\quad \tilde\bSigma_2=\frac{1}{Tp_1p_2}\sum_{t=1}^T\bQ_C\bF_t^\prime\bR^\prime\bR\bF_t\bQ_C^\prime,
 \]
 and the eigenvalue decompositions
 \[
 \tilde\bSigma_1=\tilde\bGamma_1\tilde\bLambda_1\tilde\bGamma_1^\prime, \quad \tilde\bSigma_2=\tilde\bGamma_2\tilde\bLambda_2\tilde\bGamma_2^\prime.
 \]
 Then, the normalized loading and factor score matrices are
 \[
 \tilde\bR=p_1^{1/2}\bU_R\tilde\bGamma_1,\quad  \tilde\bC=p_2^{1/2}\bU_C\tilde\bGamma_2,\quad \tilde\bF_t=\tilde\bGamma_1^\top\bQ_R\bF_t\bQ_C^\prime\tilde\bGamma_2.
 \]
We are actually estimating $\tilde\bR$, $\tilde\bC$ and $\tilde\bF_t$. Moreover, in the iterative algorithm, we will normalize the estimators similarly in each step, so that condition (\ref{id}) is always satisfied.

The simulation results for $\tau\ne 0.5$ and the results for dependent idiosyncratic errors are postponed to the Supplementary Material to save space and comply with the page limit.

\subsection{Determining the numbers of factors for $\tau=0.5$}
This section aims to verify the effectiveness of the  proposed methods for estimating the numbers of row and column factors, when $\tau=0.5$. Table \ref{tab:freq} reports the frequencies of exact estimation  when $(T,p_1,p_2)$ grows gradually and the noises are sampled from different distributions.
The approaches proposed in \cite{Chen2023Statistical} and \cite{yu2022projection} are taken as competitors, which are also designed for matrix factor models.  Another natural idea is to first vectorize the data matrices $\bX_t$ and then use the approach in \cite{Chen2021quantile}, which expects to lead to an estimation of the total number of $k=k_1k_2$ factors in theory. $K_1, K_2$ are set as 6 for matrix factor models while $k_{\max}=12$ for \cite{Chen2021quantile}'s method. The $\theta^*$ is set to be 3.

Following \cite{Chen2021quantile}, for rank-minimization we set $C_{p_1p_2T}=\delta L_{p_1p_2T}^{2/3}$, where $\delta=(\hat\sigma_{T,1}^{K_1}+\hat\sigma_{T,1}^{K_2})/2$. For the information criterion, we actually use an accelerated algorithm in the simulation rather than direct grid search in $\{1,...,K_1\}\times\{1,...,K_2\}$. In detail, we first fix $l_2=K_2$ and estimate $k_1$ by grid search in $\{1,...,K_1\}$. Next, we fix $l_1=\hat k_1$ and  estimate $k_2$ by grid search in $\{1,...,K_2\}$. The thresholding parameter for the information criterion is set as $C_{p_1p_2T}=\delta L_{p_1p_2T}$, which is slightly smaller than that for rank-minimization.

By Table \ref{tab:freq}, when the noises are from the standard normal distribution, the proposed three approaches  with matrix quantile factor model perform comparably with the $\alpha$-PCA  ($\alpha=0$) by \cite{Chen2023Statistical} and the projected estimation (PE)  by \cite{yu2022projection}. On the other hand, when the noises are from heavy-tailed distributions $t_3$ or $t_1$, the $\alpha$-PCA and PE methods gradually lose accuracy, while the proposed three methods remain reliable, due to the robustness of check loss functions. The vectorized method doesn't work in this example mainly because the dimensions are much smaller compared with the settings in \cite{Chen2021quantile} and we are considering weak signals with large $\theta^*$. Moreover, the data matrix after vectorization is severely unbalanced ($p_1p_2\gg T$), making the idiosyncratic errors matter too much.

\begin{table}[ht]
\centering
			{\fontsize{12}{11.5}\selectfont
	\caption{The frequencies of exactly estimating $( k_1, k_2)$ (or $k_1\times k_2$ for the vectorized model) by different approaches over 500 replications. \label{tab:freq}}
		\begin{tabular}{cccccccccccccccc}
			\hline
			$\bE_t$&$T$&$p_1=p_2$
			 &mqf-ER&mqf-RM&mqf-IC&$\alpha$-PCA&PE&vqf-RM\\\hline
		 \multirow{9}{*}{$\mathcal{N}(0,1)$}&20&20&0.83&0.89&0.00&0.68&0.83&0.03
\\
		 &20&50&0.99&0.92&0.52&0.98&0.99&0.18
\\
		 &20&80&1.00&0.98&0.89&1.00&1.00&0.21
\\
		 &50&20&1.00&1.00&0.00&0.98&1.00&0.26
\\
		 &50&50&1.00&0.98&1.00&1.00&1.00&0.03
\\
		 &50&80&1.00&1.00&1.00&1.00&1.00&0.07
\\
		 &80&20&1.00&1.00&0.00&1.00&1.00&0.53
\\
		 &80&50&1.00&1.00&1.00&1.00&1.00&0.02
\\
		 &80&80&1.00&1.00&1.00&1.00&1.00&0.02
\\\hline
		 \multirow{9}{*}{$t_3$}&20&20&0.49&0.12&0.00&0.03&0.05&0.14
\\
		 &20&50&0.97&0.68&0.32&0.27&0.31&0.15
\\
		 &20&80&0.99&0.89&0.64&0.36&0.40&0.16
\\
		 &50&20&0.90&0.31&0.00&0.06&0.12&0.15
\\
		 &50&50&1.00&0.90&1.00&0.39&0.45&0.12
\\
		 &50&80&1.00&0.99&1.00&0.69&0.63&0.07
\\
		 &80&20&0.98&0.16&0.00&0.14&0.23&0.03
\\
		 &80&50&1.00&1.00&1.00&0.76&0.68&0.02
\\
		 &80&80&1.00&1.00&1.00&0.86&0.77&0.01
\\\hline
		 \multirow{9}{*}{$t_1$}&20&20&0.09&0.00&0.00&0.01&0.00&0.01
\\
		 &20&50&0.83&0.40&0.06&0.01&0.00&0.00
\\
		 &20&80&0.97&0.63&0.67&0.01&0.00&0.01
\\
		 &50&20&0.60&0.04&0.00&0.03&0.01&0.01
\\
		 &50&50&1.00&0.53&0.97&0.02&0.01&0.01
\\
		 &50&80&1.00&0.89&1.00&0.00&0.00&0.00
\\
		 &80&20&0.68&0.00&0.00&0.03&0.02&0.01
\\
		 &80&50&1.00&0.85&1.00&0.01&0.00&0.01
\\
		 &80&80&1.00&0.98&1.00&0.00&0.00&0.01\\\hline
	\end{tabular}
}
\end{table}

\subsection{Estimating loadings and factor scores for $\tau=0.5$}
Next, we investigate the accuracy of the estimated loadings and factor scores by different approaches. We use the similar settings as in Table \ref{tab:freq} and let $\tau=0.5$. Note that the minimizers to the check loss function are not unique, so the estimated loading matrices  converge only after a rotation. Due to such an identification issue, we will mainly focus on the estimation accuracy of the loading spaces. Let $\bR_0$ and $\hat \bR$ be the true and estimated loading matrices respectively, both satisfying the identification condition in (\ref{id}). We define the distance between the two loading spaces by
\[
\mathcal{D}(\bR_0,\hat\bR)=\bigg(1-\frac{1}{k_1p_1^2}\text{tr}(\hat\bR^\prime\bR_0\bR_0^\prime\hat\bR)\bigg)^{1/2}.
\]
It's easy to see that $\mathcal{D}(\bR_0,\hat\bR)$ always takes value in the interval $[0,1]$. A smaller value of $\mathcal{D}(\bR_0,\hat\bR)$ indicates more accurate estimation of $\bR_0$. When $\mathcal{D}(\bR_0,\hat\bR)=0$, the two loading spaces are exactly the same. Similar distance can be defined between $\bC_0$ and $\hat \bC$. Let $\bW_0=\bC_0\otimes\bR_0$, $p=p_1p_2$ and $\hat\bW$ be an estimate of $\bW_0$. Similarly, we define
$$
\mathcal{D}(\bW_0,\hat\bW)=\bigg(1-\frac{1}{kp^2}\text{tr}(\hat\bW^\prime\bW_0\bW_0^\prime\hat\bW)\bigg)^{1/2}.
$$
The existing vector quantile factor analysis estimates $\bW_0$ by $\hat\bW=\hat\bL$ given in \cite{Chen2021quantile}. The matrix quantile factor analysis estimates $\bW_0$ by the plug-in estimator $\hat\bW=\hat{\bC}\otimes\hat{\bR}$.

Table \ref{tab:loading} reports the estimation accuracy of the loading spaces by different methods over 500 replications. The conclusions almost follow those in \ref{tab:freq}. The estimation based on matrix quantile factor models (``mqf'') is accurate and stable  under all settings, while $\alpha$-PCA and ``PE'' only work for light-tailed cases.  Even under the normal cases, ``mqf'' can outperform ``PE'', mainly because the latter only contains one-step iteration thus relying on a good initial projection direction. There are some enormous errors for $\alpha$-PCA, ``PE'' and the vectorized method in the table.

 \begin{table}[ht]
 	\centering
			{\fontsize{11.5}{10.5}\selectfont
 	\caption{ Distances between the estimated loading space and the truth by different approaches over 500 replications. ``mqf'' stands for matrix quantile factor analysis, while ``vqf'' stands for vectorized quantile factor analysis.  \label{tab:loading} }
 		\begin{tabular}{cccccccccccccccc}
 			\hline
 			 \multirow{2}{*}{$\bE_t$}&\multirow{2}{*}{$T$}&\multirow{2}{*}{$p_1=p_2$}&\multicolumn{3}{l}
 {$\mathcal{D}(\bR_0,\hat\bR)$}&\multicolumn{3}{l}{$\mathcal{D}(\bC_0,\hat\bC)$}&\multicolumn{2}{l}
 {$\mathcal{D}(\bW_0,\hat{\bW})$}\\
 			 &&&mqf&$\alpha$-PCA&PE&mqf&$\alpha$-PCA&PE&mqf&vqf\\\hline
 		 \multirow{9}{*}{$\mathcal{N}(0,1)$}&20&20&0.04&0.09&0.08&0.05&0.11&0.09&0.06&0.69
\\
 		 &20&50&0.02&0.06&0.05&0.03&0.08&0.07&0.04&0.74
\\
 		 &20&80&0.02&0.04&0.04&0.03&0.06&0.05&0.03&0.75
\\
 		 &50&20&0.02&0.04&0.04&0.02&0.06&0.05&0.03&0.36
\\
 		 &50&50&0.01&0.04&0.04&0.02&0.05&0.05&0.02&0.50
\\
 		 &50&80&0.01&0.03&0.02&0.01&0.03&0.03&0.02&0.38
\\
 		 &80&20&0.01&0.03&0.03&0.02&0.04&0.04&0.02&0.17
\\
 		 &80&50&0.01&0.03&0.03&0.01&0.04&0.03&0.02&0.25
\\
 		 &80&80&0.01&0.02&0.02&0.01&0.03&0.03&0.01&0.22
\\\hline
 		 \multirow{9}{*}{$t_3$}&20&20&0.06&0.50&0.50&0.07&0.53&0.43&0.09&0.85
\\
 		 &20&50&0.03&0.23&0.23&0.04&0.34&0.21&0.05&0.85
\\
 		 &20&80&0.02&0.17&0.17&0.03&0.26&0.15&0.04&0.85
\\
 		 &50&20&0.03&0.33&0.31&0.04&0.46&0.29&0.05&0.66
\\
 		 &50&50&0.02&0.20&0.18&0.02&0.27&0.15&0.03&0.66
\\
 		 &50&80&0.01&0.10&0.10&0.02&0.16&0.09&0.02&0.61
\\
 		 &80&20&0.02&0.33&0.30&0.03&0.43&0.27&0.04&0.52
\\
 		 &80&50&0.01&0.10&0.08&0.01&0.15&0.08&0.02&0.32
\\
 		 &80&80&0.01&0.07&0.06&0.01&0.09&0.06&0.01&0.29
\\\hline
 		 \multirow{9}{*}{$t_1$}&20&20&0.08&0.95&0.95&0.10&0.92&0.92&0.13&0.98
\\
 		 &20&50&0.03&0.98&0.98&0.05&0.97&0.97&0.06&0.97
\\
 		 &20&80&0.03&0.99&0.99&0.03&0.98&0.98&0.04&0.98
\\
 		 &50&20&0.03&0.95&0.95&0.04&0.92&0.93&0.05&0.75
\\
 		 &50&50&0.02&0.98&0.98&0.02&0.97&0.97&0.03&0.81
\\
 		 &50&80&0.01&0.99&0.99&0.02&0.98&0.98&0.02&0.80
\\
 		 &80&20&0.03&0.95&0.95&0.04&0.92&0.92&0.05&0.79
\\
 		 &80&50&0.01&0.98&0.98&0.02&0.97&0.97&0.02&0.61
\\
 		 &80&80&0.01&0.99&0.99&0.01&0.98&0.98&0.02&0.57\\\hline
  	\end{tabular}}
 \end{table}

\subsection{Asymptotic distribution}

\begin{figure}[htp]
	\centering
	\begin{minipage}[c]{4cm}
		\centering
		\includegraphics[width=4 cm]{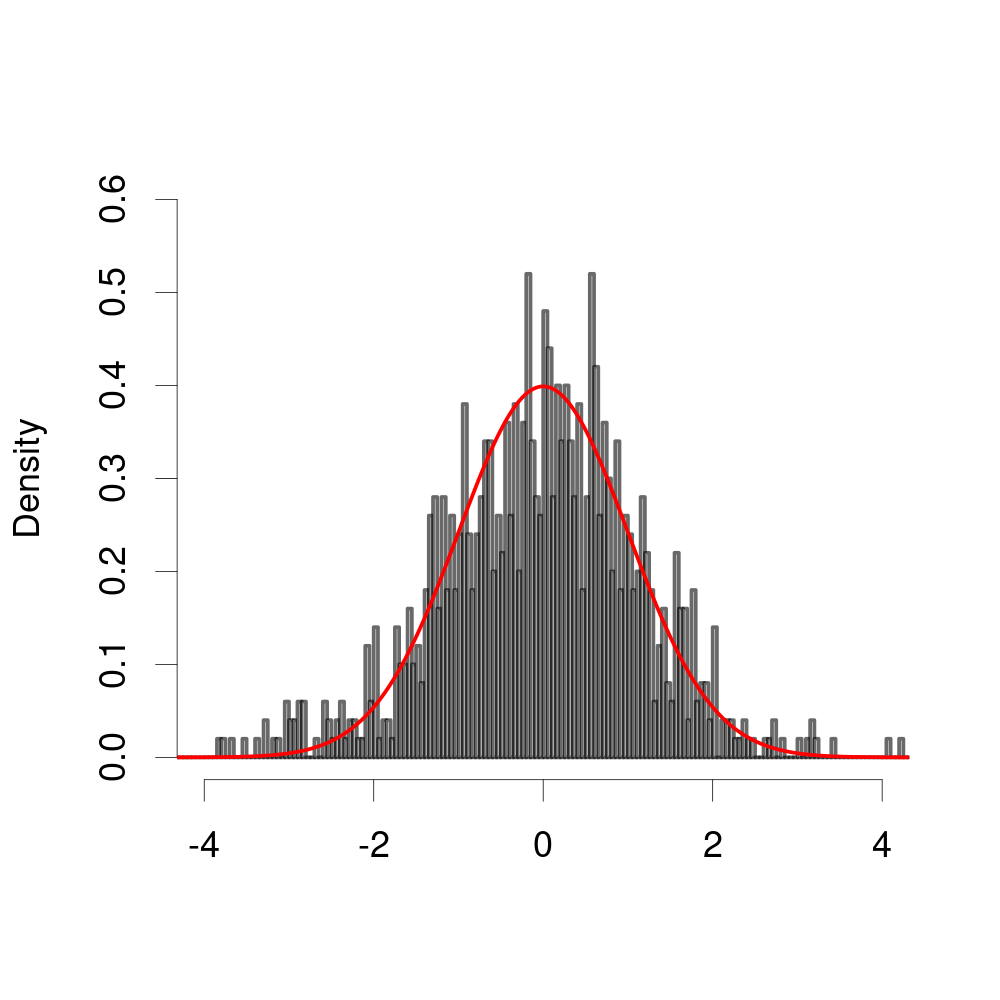}
		 \caption*{$\varepsilon_{ijt}\sim\mathcal{N}(0,1),\tau=0.5$}
	\end{minipage}
	\begin{minipage}[c]{4cm}
	\centering
	\includegraphics[width=4 cm]{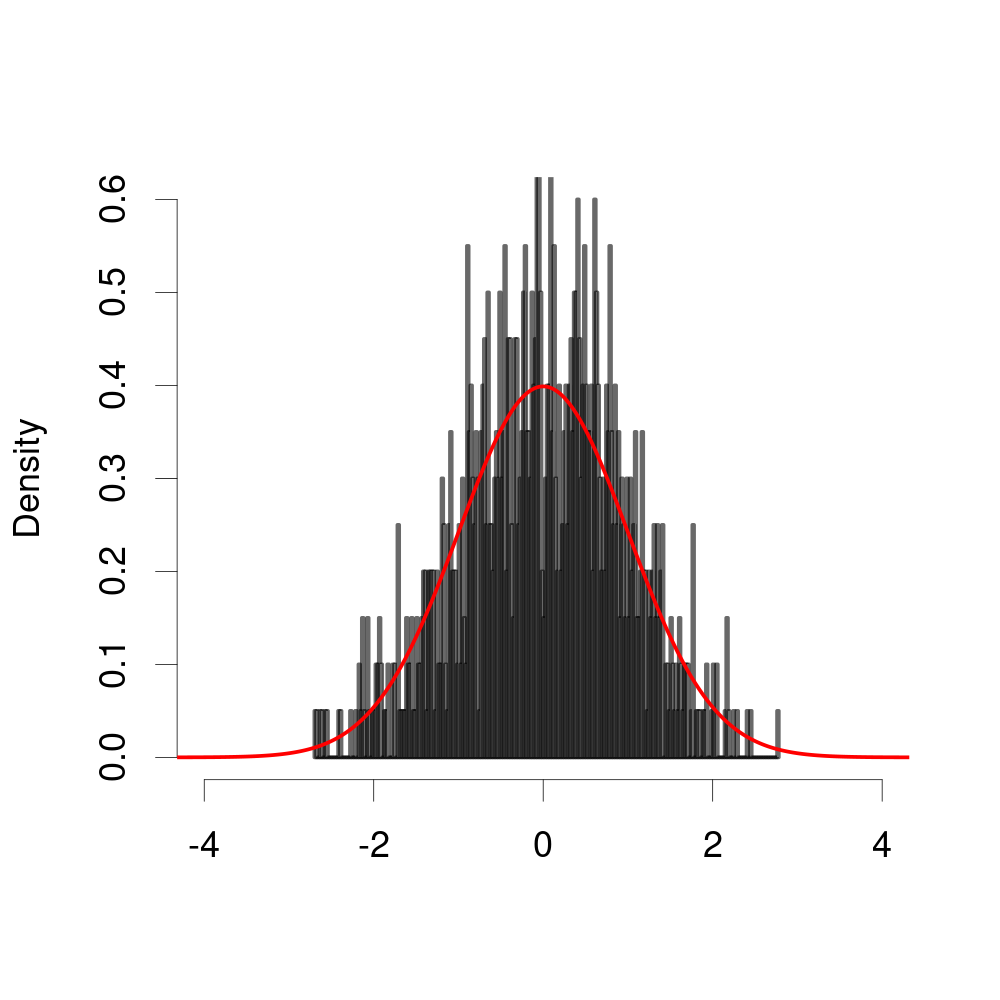}
	\caption*{$\varepsilon_{ijt}\sim t_3,\tau=0.5$}
\end{minipage}
	\begin{minipage}[c]{4cm}
	\centering
	\includegraphics[width=4 cm]{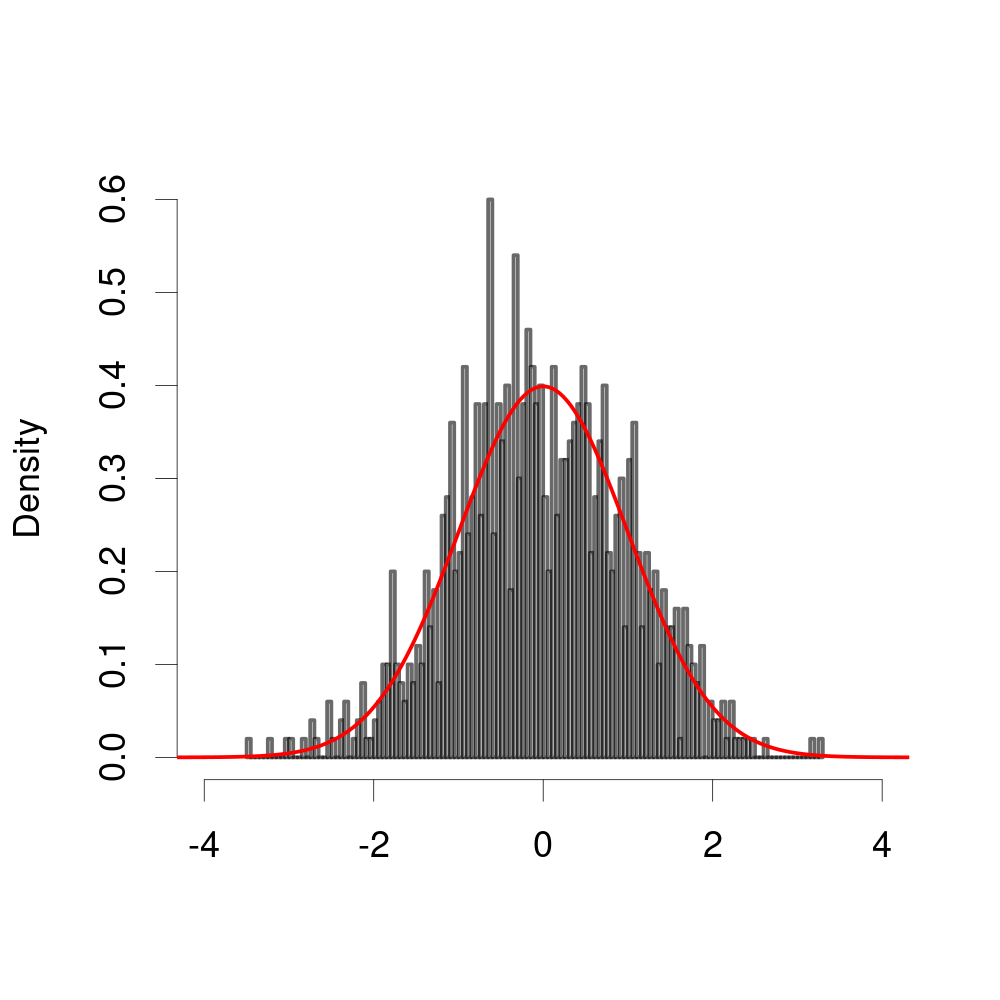}
	\caption*{$\varepsilon_{ijt}\sim t_1,\tau=0.5$}
\end{minipage}
\caption{Empirical densities of $\hat\bR_{11}$ after standardization under various settings with $T=p_1=p_2=50$. }\label{fig:dis}
\end{figure}

We verify the asymptotic distributions of the smoothly estimated factor loadings in Theorem \ref{CLT}. We set $\theta^*=g_t=1$ in (\ref{data generating}) and generate $\bF_t$ from i.i.d. $\mathcal{N}(0,1)$. Figure \ref{fig:dis} plots the empirical density of $\hat \bR_{11}$ after standardization according to Theorem \ref{CLT}, over 1000 replications with $\tau=0.5$ when the entries of $\bE_t$ are i.i.d. from $\mathcal{N}(0,1)$, $t_3$ or $t_1$. Figure \ref{fig:dis} clearly shows the asymptotic normality of the estimators with well fitted variances.

\section{Real Data Analysis}\label{RD}
\subsection{Data description}
In this section, we apply the proposed matrix quantile factor model and associated estimators to the analysis of a real data set, the Fama-French 100 portfolios data set. This an open resource provided by Kenneth R. French, which can be downloaded from the website \url{http://mba.tuck.dartmouth.edu/pages/faculty/ken.french/data_library.html}. It contains monthly return series of 100 portfolios, structured into a $10\times 10$ matrix according to 10 levels of market capital size (S1-S10) and 10 levels of book-to-equity ratio (BE1-BE10). Considering the missing rate, we use the data from 1964-01 to 2021-12 in this study, covering 696 months. Similar data set has ever  been studied in \cite{wang2019factor} and \cite{yu2022projection}. The data library also provides information on Fama-French three factors and excess market returns.  Following the preprocessing steps in \cite{wang2019factor} and \cite{yu2022projection}, we first subtract the excess market returns and standardize each of the portfolio return series. In the first step, we provide some descriptive information of the data set in the Supplementary Material to save space.

\subsection{Estimation}
In the second step, we fit the matrix quantile factor model. The numbers of row and column factors should be determined first. Table \ref{tab: ff tau} provides the estimated $(k_1,k_2)$ using the proposed three approaches at different quantiles $\tau$. The results by the vectorized method with \cite{Chen2021quantile} are also reported in the table, which leads to the estimation of total number of factors. By Table \ref{tab: ff tau}, the proposed eigenvalue ratio method and information criterion always lead to an estimate of $\hat k_1=\hat k_2=1$, while the rank minimization approach gives more row and/or column factors when $\tau\in [0.15,0.9]$. The vectorized method leads to an estimate of 2 factors in total at most quantiles.  Based on the results,  there should be at least one powerful row factor and column factor in the system, and potentially one weak row factor and/or column factor. When $\tau$ is at the edge, the leading factor becomes more influential. On the other hand, the approaches in \cite{wang2019factor} and \cite{yu2022projection} will both lead to $\hat k_1=\hat k_2=1$.  In this example, it might be a good choice to use  $\hat k_1=\hat k_2=2$ when $\tau$ is around $0.5$, and $\hat k_1=\hat k_2=1$ when $\tau$ is at the edge.

\begin{table}[htp]
 	\centering
			{\fontsize{11.5}{12}\selectfont
	\caption{Estimating results of factor numbers and loading spaces for Fama-French 100 portfolio data set at different $\tau$. $S^R_{12}$ is an abbreviation for $S(\hat\bR_1,\hat\bR_2)$.\label{tab: ff tau}}
	%\resizebox{\columnwidth}{!}{
		\begin{tabular}{lllllllllllllllllll}
			\hline
			 \multirow{2}{*}{$\tau$}&\multicolumn{3}{l}{$(\hat k_1,\hat k_2)$}&$\widehat{k_1\times k_2}$&\multicolumn{4}{l}{Similarity of loading spaces}\\\cmidrule(lr){2-4}\cmidrule(lr){6-9}
			 &mqf-ER&mqf-RM&mqf-IC&vqf-RM&$S^R_{mqf,PE}$&$S^C_{mqf,PE}$&$S^{RC}_{mqf,vqf}$&$S^{RC}_{PE,vqf}$\\\hline
			 0.05&(1,1)&(1,1)&(1,1)&1&0.503&0.661&0.381&0.691
			\\
			 0.1&(1,1)&(1,1)&(1,1)&1&0.965&0.914&0.736&0.725
			\\
			 0.15&(1,1)&(1,2)&(1,1)&2&0.992&0.961&0.723&0.719
			\\
			 0.2&(1,1)&(2,1)&(1,1)&2&0.991&0.957&0.724&0.716
			\\
			 0.25&(1,1)&(2,2)&(1,1)&2&0.998&0.979&0.737&0.726
			\\
			 0.3&(1,1)&(2,2)&(1,1)&2&0.996&0.991&0.747&0.744
			\\
			 0.35&(1,1)&(2,2)&(1,1)&2&0.997&0.996&0.738&0.737
			\\
			 0.4&(1,1)&(2,2)&(1,1)&2&0.998&0.996&0.722&0.723
			\\
			 0.45&(1,1)&(2,2)&(1,1)&2&0.997&0.992&0.707&0.702
			\\
			 0.5&(1,1)&(2,2)&(1,1)&2&0.999&0.993&0.674&0.666
			\\
			 0.55&(1,1)&(2,2)&(1,1)&2&0.988&0.996&0.673&0.659
			\\
			 0.6&(1,1)&(2,2)&(1,1)&2&0.993&0.997&0.674&0.676
			\\
			 0.65&(1,1)&(2,2)&(1,1)&2&0.989&0.998&0.705&0.703
			\\
			 0.7&(1,1)&(2,2)&(1,1)&2&0.990&0.994&0.740&0.737
			\\
			 0.75&(1,1)&(2,2)&(1,1)&2&0.975&0.982&0.727&0.716
			\\
			 0.8&(1,1)&(1,2)&(1,1)&2&0.977&0.924&0.743&0.727
			\\
			 0.85&(1,1)&(1,2)&(1,1)&2&0.958&0.967&0.723&0.703
			\\
			 0.9&(1,1)&(2,2)&(1,1)&1&0.985&0.937&0.711&0.699
			\\
			 0.95&(1,1)&(1,1)&(1,1)&1&0.884&0.873&0.607&0.627\\\hline
	\end{tabular}}
\end{table}

The next step is to estimate the loading matrices and factor scores with $\hat k_1=\hat k_2=2$.
It's worth noting that the quantile factor models can handle missing values naturally by optimization only with non-missing entries, e.g., in the completely random missing case by defining
\[
\mathbb{M}_{p_1p_2T}(\theta)=\frac{1}{p_1p_2T}\sum_{(i,j,t)\in \mathcal{M}}\rho_{\tau}(X_{ijt}-\mathbf{r}_i^\prime\bF_t \mathbf{c}_j),
\]
where $\mathcal{M}$ indicates the index set of all non-missing entries. However, the $\alpha$-PCA and ``PE'' methods require to impute the missing entries first. Considering that the missing rate is small in this example ($0.23\%$), we use simple linear interpolation method to impute missing data. To measure the similarity of two estimated loading spaces, we define the following indicator:
\[
S(\hat \bR_1,\hat \bR_2)=\frac{1}{k_1}\text{tr}\bigg(\frac{1}{p_1^2}\hat\bR_1^\prime\hat\bR_2\hat\bR_2^\prime\hat\bR_1\bigg),
\]
where $\hat \bR_1$ and $\hat\bR_2$ are the estimated $p_1\times k_1$ row loading matrices. Note that the columns of $\hat \bR_1$ and $\hat\bR_2$ are orthogonal after scaling. Therefore, the value $p_1^{-1}\hat\bR_2\hat\bR_2^\prime\hat\bR_1$ is actually the projection matrix of $\hat\bR_1$ to the space of $\hat\bR_2$. The value of $S(\hat \bR_1,\hat \bR_2)$ will always be in the interval $[0,1]$. When the two loading spaces are closer to each other, the value of $S(\hat \bR_1,\hat \bR_2)$ will be larger.

The last four columns of Table \ref{tab: ff tau} report the similarity of estimated loading spaces by matrix-quantile-factor-model and two competitors,  ``PE'' and the vectorization approach. For the vectorization, we calculate similarity by considering the Kronecker product $\hat\bC\otimes \hat\bR$. It's seen that the similarity indicators for the matrix-quantile-factor based approach and ``PE'' approach are very close to 1,  implying that the estimated loading spaces are almost the same,  especially when $\tau$ is near 0.5. However, when $\tau$ is at the edge, the difference of the estimated loading spaces becomes more significantly.  For the vectorization approach, the estimated loading space is always not similar to that from the matrix models, consistent with our findings from the simulation study.

\subsection{Interpretation}\label{interp}
Now we aim to interpret the matrix quantile factors in this example.
Table \ref{tab: ff loading} presents the estimated $\hat \bR$ and $\hat\bC$ by matrix quantile factor model at $\tau=0.5$, $0.05$, $0.95$ as well as those by ``PE`''.  By Table \ref{tab: ff loading}, the effects of the row factors and column factors are closely related to market capital sizes and book-to-equity ratios. From the perspective of size ($\hat\bC$), under the matrix quantile factor model with $\tau=0.5$, the small-size portfolios load more heavily on the first factor than the large-size portfolios. Moreover, the second factor has opposite effects on small-size portfolios and large-size ones. Similar results are found for the ``PE'' method, although the values of loadings are not exactly the same. Taking $\tau$ at edge will lead to different finding, where the first factor has more significant effect on the large-size portfolios. In other words, the edge quantile factors show disparate information of the data.

From the perspective of book-to-equity ratio ($\hat\bR$), with $\tau=0.5$, the large-BE portfolios load more heavily on the first factor than small-BE ones, while the second factor has opposite effects on the two classes. The ``PE'' factors show similar trend after orthogonal transformation (changing sign). When $\tau=0.05$ and $\tau=0.95$, the first factor depends on the average of all portfolios. It's worth noting that the reported row and column factors are highly suggestive, because they coincide with financial theories. The capital size and book-to-equity ratio are known to be two important factors affecting portfolio returns in negative collaboration. The row and column factors in this example might be closely related to the SMB and HML factors in portfolio theory.

We also verify the robustness of the check losss function by the matrix quantile factor model
with this real example. The results are postponed in the Supplementary Material.

\begin{table*}[hbpt]
 	\centering
			{\fontsize{11.47}{12}\selectfont
		\caption{ Transposed loading matrices for Fama-French data set by matrix quantile factor model at $\tau=0.5$. }\label{tab: ff loading}
		\renewcommand{\arraystretch}{1}
		%\scalebox{1}{
		 \begin{tabular*}{16cm}{cc|ccccccccccc}
				\hline
				\multicolumn{13}{l}{Size ($\hat\bC$)}\\
				\hline
				 Methods&Factors&S1&S2&S3&S4&S5&S6&S7&S8&S9&S10\\\hline
				 \multirow{2}{*}{$mqf_{0.5}$}&1&1.15&1.20&1.25&1.21&1.14&1.08&0.90&0.78&0.55&0.00
				\\
				 &2&1.28&0.83&0.53&0.23&-0.22&-0.57&-0.86&-1.11&-1.66&-1.49\\
				\hline
				 \multirow{2}{*}{PE}&1&-1.17&-1.21&-1.26&-1.20&-1.15&-1.04&-0.90&-0.81&-0.54&0.01\\
				 &2&1.39&0.95&0.48&0.18&-0.28&-0.65&-0.91&-1.18&-1.56&-1.31
				\\\hline
				 \multirow{2}{*}{$mqf_{0.05}$}&1&0.81&0.87&0.85&0.99&1.02&1.07&1.03&1.05&1.12&1.14
				\\
				 &2&1.04&1.30&1.42&0.66&0.68&-0.18&-0.54&-0.85&-1.30&-1.25\\
				\hline
				 \multirow{2}{*}{$mqf_{0.95}$}&1&0.83&0.84&0.86&0.87&0.97&1.04&1.03&1.05&1.15&1.28
				\\
				 &2&-0.63&0.15&0.71&0.51&0.04&0.83&-0.82&-0.83&-1.87&1.81\\\hline
				\multicolumn{13}{l}{Book-to-Equity ($\hat\bR$)}\\\hline
				 Methods&Factors&BE1&BE2&BE3&BE4&BE5&BE6&BE7&BE8&BE9&BE10\\\hline
				 \multirow{2}{*}{$mqf_{0.5}$}&1&0.59&0.74&0.95&1.05&1.12&1.16&1.12&1.13&1.08&0.89
				\\
				 &2&2.13&1.67&0.75&0.29&-0.07&-0.58&-0.68&-0.73&-0.67&-0.52\\
				\hline
				 \multirow{2}{*}{PE}&1&-0.53&-0.83&-1.01&-1.08&-1.10&-1.12&-1.12&-1.08&-1.07&-0.90
				\\
				 &2&2.16&1.56&0.81&0.31&-0.15&-0.50&-0.69&-0.77&-0.70&-0.55
				\\\hline
				 \multirow{2}{*}{$mqf_{0.05}$}&1&0.98&0.90&0.94&1.09&1.11&1.07&0.96&1.03&0.86&1.05
				\\
				 &2&1.33&1.37&1.20&0.61&0.08&-0.55&-0.81&-1.04&-1.40&-0.74\\\hline
				 \multirow{2}{*}{$mqf_{0.95}$}&1&0.99&1.02&1.01&1.00&1.00&1.01&0.99&1.00&0.99&0.98
				\\
				 &2&-1.61&0.39&1.84&-0.17&-0.09&-1.44&0.00&0.26&-0.42&1.23
				\\ 				\hline	
		\end{tabular*}}		
\end{table*}

\subsection{Usefulness in prediction}
By Table \ref{tab: ff tau}, when $\tau$ is at the edge, the similarity indicator decreases, suggesting that considering the edge quantiles might be helpful for extracting extra information from the data. However, by Figure 3 in the supplementary material, %\ref{fig:similarity},
 the low similarity can also potentially results from the reduced stability. Therefore, to justify the usefulness of the proposed model, we construct a rolling prediction procedure as follows. Let $y_t$ be any of the  Fama-French three factors at month $t$. We consider a forecasting model for $y_t$:
\[
y_{t+1}=\alpha+\beta y_{t}+\gamma^\prime \bF_{t+1}+e_{t+1},
\]
where $\bF_t$ is a vector of estimated  factors from the Fama-French 100 portfolio data set.  We estimate $\alpha,\beta,\gamma$ using ordinary least squares.  For $\bF_t$, we consider eight specifications: (i) $\bF_t=0$, which is the benchmark AR(1) model, (ii) $\bF_t$ from ``PE'', (iii) $\bF_t$ from ``PE'' and matrix quantile factor model at $\tau=0.05$, (iv)
$\bF_t$ from ``PE'' and matrix quantile factor model at $\tau=0.95$, (v) $\bF_t$ from ``PE'' and matrix quantile factor model at $\tau=0.05$ and $\tau=0.95$, (vi) to (viii) generate $\bF_t$ similarly to (iii) to (v) but replacing matrix quantile factors with vectorized quantile factors. To control the dimension of the design matrix, we use $k_1=k_2=1$ in this part, and ignore the case $\tau=0.5$ because the estimated loading space is very close to that from ``PE'' by Table \ref{tab: ff tau}. To predict $y_{t+1}$, we first estimate all the factors using  historical data before (inclusive) $(t+1)$ with a  rolling window  of 60 months, and then fit the predicting model using data only before $(t+1)$. The predictor $\hat y_{t+1}$ then follows the fitted model. Table \ref{tab: ff prediction} reports the root of mean squared error (RMSE) and mean absolute error (MAE) for the prediction over all the periods, from different predicting models and for different  Fama-French factors. As shown in the table, adding estimated factors into the model helps reduce the error for the SMB factor and the HML factor, while considering the edge quantiles further improves the prediction performance. This is consistent with our interpretation in Section \ref{interp}. The estimated row and column factors from the matrix quantile factor model are closely related to the Fama and French SMB and HML factors. In this example, the matrix quantile factor model leads to smaller MAE while the vectorized model leads to smaller RMSE. But for the RF factor, the Benchmark AR(1) model works already the best. Adding more factors into the predictors only results in more errors, mainly because the market excess return has already been removed from the data.

\begin{table*}[hbpt]
 	\centering
			{\fontsize{11.5}{12}\selectfont
		\caption{RMSE and MAE from the rolling prediction procedure, by different models and for different Fama-French factors.}\label{tab: ff prediction}
		\renewcommand{\arraystretch}{1}
	%	\scalebox{1}{ 	
	 \begin{tabular*}{16cm}{lllllllllllll}
				\hline
				&\multicolumn{2}{l}{SMB factor}&\multicolumn{2}{l}{HML factor}&\multicolumn{2}{l}{RF factor}\\\cmidrule(lr){2-3}\cmidrule(lr){4-5}\cmidrule(lr){6-7}
				Predicting models&RMSE&MAE&RMSE&MAE&RMSE&MAE\\\hline
				AR benchmark&3.182&2.272&3.020&2.185&0.065&0.040
				\\
				AR plus $\hat \bF_{PE}$&1.692&1.101&2.953&2.180&0.065&0.041
				\\\hline
				AR plus $\hat \bF_{PE}$, $\hat\bF_{mqf}^{\tau=0.05}$&1.610&1.058&3.045&2.199&0.066&0.041
				\\
				AR plus $\hat \bF_{PE}$, $\hat\bF_{mqf}^{\tau=0.95}$&1.563&1.032&2.975&2.190&0.065&0.042
				\\
				AR plus $\hat \bF_{PE}$, $\hat\bF_{mqf}^{\tau=0.05}$, $\hat\bF_{mqf}^{\tau=0.95}$&1.441&0.924&2.824&2.057&0.067&0.042
				\\\hline
				AR plus $\hat \bF_{PE}$, $\hat\bF_{vqf}^{\tau=0.05}$&1.624&1.063&3.019&2.195&0.066&0.041
				\\
				AR plus $\hat \bF_{PE}$, $\hat\bF_{vqf}^{\tau=0.95}$&1.589&1.039&2.976&2.178&0.065&0.041
				\\
				AR plus $\hat \bF_{PE}$, $\hat\bF_{vqf}^{\tau=0.05}$, $\hat\bF_{vqf}^{\tau=0.95}$&1.429&0.932&2.815&2.069&0.066&0.042\\\hline
		\end{tabular*}}		
\end{table*}

\subsection{Usefulness in imputing missing values}
\begin{figure}[htp]
	\centering
	\includegraphics[width=14cm]{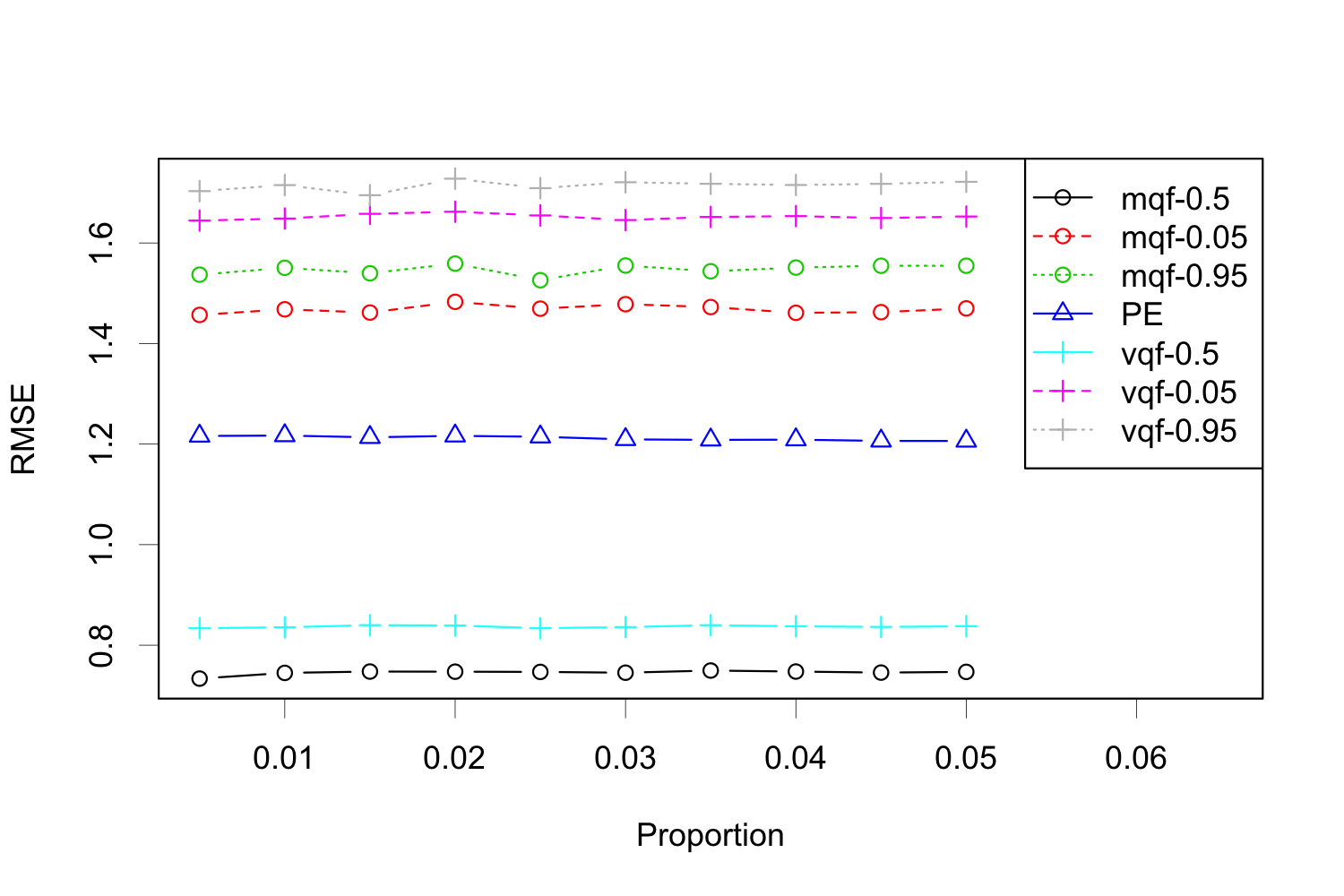}
	\caption{Mean imputing error (RMSE) as the kicking-out proportion increases.}
	\label{fig:missing}
\end{figure}

In our last experiment, we investigate the performance of matrix quantile factor model in imputing missing entries. We deliberately kick out a proportion of entries from the data, and treat them as missing values. Then, we fit a factor model, estimate the loading and factor score matrices, and impute the missing entries by the estimated common components. We calculate the imputing error in terms of RMSE, denoted by $a_1$. As a benchmark, we also calculate the imputing error when simply imputing the missing values with 0 (the data are standardized), denoted as $a_0$.
For robustness check, we repeat the procedure 50 times and report the averaging $a_1/a_0$ under different factor models in Figure \ref{fig:missing}, as the kicking-out proportion increases. It's seen that the matrix-quantile factor model with $\tau=0.5$ leads to the lowest imputing error in all scenarios.

\section{Conclusion and Discussion}\label{CD}

In this study, we proposed a matrix quantile factor model that is a hybrid of quantile feature representation and a low rank structure for matrix data. By minimizing the check loss function under rotation constraints, we obtain estimates of the row and column factor spaces that are proved to be consistent in the sense of Frobenious norm. Three model selection criteria were given to consistently determine simultaneously the numbers of row and column factors. Central limit theorems are derived for the smoothed loading estimates by novelly introducing an equivalent augmented Lagrangian function. There are at least three problems that are worthy of being studied in the future. First, a statistical test for the presence of the low-rank matrix structure in the matrix quantile factor model is of potential usefulness as a model checking tool. Second, the latent factor structure here can be extended to the case where both observable explanatory variables and latent factors are incorporated into modeling the quantiles of matrix sequences. Third, the computation error with the algorithm, that parallels to the statistical error given in our theorem, is still unknown. We leave all these to our future research work.

\section*{Supplementary Material}
The Supplementary Material contains extra numerical results and real data analysis, the detailed technical proof of the main theorems, as well as some technical lemmas that are of their own interests.

%\section*{Acknowledgements}
%Kong's work is partially supported by NSF China (72342019). Liu's work is partially supported by NSF China (12001278) and Natural Science Research of Jiangsu Higher Education Institutions of China (20KJB110017). Yu's research is partially supported by NSF China (12301350), Shanghai Pujiang Program (23PJ1402700), and the Fundamental Research Funds for the Central Universities, China.
%Zhao's work is partially supported by NSF China (11871252) and A Project Funded by the Priority Academic Program Development of Jiangsu Higher Education Institutions, China.

\bibliographystyle{chicago}
\bibliography{Ref}
\end{document}